\documentstyle{article}

\newcommand{\ba}{\begin{eqnarray}}
\newcommand{\ea}{\end{eqnarray}}
\begin{document}
\pagestyle{plain}

\title{On the relation between algebraic and configuration space 
calculations of molecular vibrations}

\author{ F. P\'erez-Bernal$^{1)}$, R. Bijker$^{2)}$, A. Frank$^{2,3)}$, 
R. Lemus$^{2)}$ and J.M. Arias$^{1)}$\\
\and
\begin{tabular}{rl}
$^{1)}$ & Departamento de F\'{\i}sica At\'omica, Molecular y Nuclear,\\
        & Facultad de F\'{\i}sica, Universidad de Sevilla,\\
        & Apdo. 1065, 41080 Sevilla, Espa\~na\\
$^{2)}$ & Instituto de Ciencias Nucleares, U.N.A.M.,\\
        & A.P. 70-543, 04510 M\'exico D.F., M\'exico\\
$^{3)}$ & Instituto de F\'{\i}sica, Laboratorio de Cuernavaca,\\
        & A.P. 139-B, Cuernavaca, Morelos, M\'exico
\end{tabular} }

\date{}
\maketitle
\noindent
\vspace{6pt}
\begin{abstract}
The relation between algebraic and traditional calculations 
of molecular vibrations is investigated. An explicit connection 
between interactions in configuration space and the corresponding 
algebraic interactions is established.
\end{abstract}
\begin{center}
\end{center}

\newpage

Ab initio calculations for rovibrational spectra of molecular 
systems attempt exact solutions of the Schr\"odinger equation. 
In practice, the molecular Hamiltonian is usually parametrized as 
a function of internal coordinates \cite{WDC} and the potential 
is modeled in terms
of force-field constants, which are determined through 
calculations involving several configurations associated to the 
molecular electronic states \cite{Raynes}. For small molecules 
this procedure is still feasible, but this is in 
general not the case for polyatomic molecules, due to the large size 
of the configuration space. It is thus important to develop 
alternative methods to describe these systems. Algebraic (or vibron) 
models 
attempt to provide such alternative techniques \cite{vibron,thebook}. 
In its original formulation \cite{vibron1,vibron2} rotations and 
vibrations were treated simultaneously in terms of coupled $U(4)$ 
algebras: ${\cal G}=U_1(4) \otimes U_2(4) \otimes \ldots$~.
For polyatomic molecules it was found to be more convenient to first 
separate the rotations and vibrations and subsequently to treat the 
vibrations in terms of coupled $U(2)$ algebras: 
${\cal G}=U_1(2) \otimes U_2(2) \otimes \ldots$~.
In the latter version of the vibron model the calculation of matrix elements 
is greatly simplified. An additional advantage is that it is well-suited 
to incorporate the underlying discrete symmetries \cite{IOFL,IO,LF}.
The vibron Hamiltonian, however, is expressed in terms of abstract algebraic 
operators, whose connection with more traditional methods has been limited
to studies of the corresponding energy surface \cite{AL}.

The aim of this letter is to investigate the relation between algebraic 
and configuration space calculations and to establish an explicit connection 
between interaction terms in coordinate space parametrizations 
and in the more abstract algebraic space. We shall illustrate this 
connection by studying 
the Be$_4$ cluster, for which the force field parameters 
were determined in \cite{RLT} by ab initio methods. 
Our comparison is based on an analysis of tetrahedral molecules 
in terms of symmetry-adapted internal coordinates \cite{Hecht}.

In the algebraic approach each relevant interatomic interaction
is associated with a $U(2)$ algebra. In the present 
example of the Be$_4$ cluster, which has a tetrahedral shape, 
there are six $U(2)$ algebras involved: 
${\cal G} = U_1(2) \otimes U_2(2) \otimes \ldots \otimes U_6(2)$.
Each $U_i(2)$ algebra ($i=1,\ldots,6$) is generated by the set 
$\{ \hat G_i \} \equiv \{ \hat N_i, \, \hat J_{+,i}, 
\, \hat J_{-,i}, \, \hat J_{0,i} \}$, satisfying the commutation 
relations 
\ba
\, [ \hat J_{0,i}, \hat J_{\pm,i}] \;=\; \pm \hat J_{\pm,i} ~,
\hspace{1cm} 
\, [ \hat J_{+,i}, \hat J_{-,i}] \;=\; 2 \hat J_{0,i} ~,
\hspace{1cm} 
\, [ \hat N_i, \hat J_{\mu,i}] \;=\; 0 ~, \label{jmui}
\ea
with $\mu=\pm,0$. Here $\hat N_i$ is the (boson) number operator 
and the operators $\hat J_{\mu,i}$ satisfy the `angular momentum' 
commutation relations of $SU_i(2)$. 
Since $\vec{J}^2_i=\hat N_i(\hat N_i+2)/4$ we can make the identification 
$j_i=N_i/2$. The eigenvalues of $\hat J_{0,i}$ are restricted to 
$m_i \geq 0$ and can have the values 
$m_i=N_i/2, (N_i-2)/2, \ldots , 1/2$ or 0 for $N_i$ odd or even, 
respectively \cite{thebook}. The local basis states for each 
oscillator are usually written as $|N_i,v_i \rangle$, where 
$v_i=(N_i-2m_i)/2=0,1, \ldots [N_i/2]$ denotes the number of oscillator 
quanta in the $i$-th oscillator. 
Because of the tetrahedral symmetry of the Be$_4$ 
cluster $N_i=N$ for the six oscillators. 
The operators in the model are expressed in terms of the 
generators of these algebras, and the symmetry requirements of the 
Hamiltonian under the 
tetrahedral group ${\cal T}_d$  can be readily imposed \cite{LF,bible}. 
In the usual algebraic formulation different chains of subgroups of 
${\cal G}$ are considered and the Hamiltonian is built by means of 
appropriate combinations of invariant operators associated to these 
chains \cite{IOFL,IO,LF}. Here we shall follow a different approach 
which leads to a richer structure for the Hamiltonian and to a direct 
connection to configuration space interactions. To achieve this goal 
we note that the local operators $\{ \hat G_i \}$
acting on bond $i$ can be projected to any of the ${\cal T}_d$ fundamental 
irreps $\Gamma=A_1$, $E$ and  $F_2$.
Using the $\hat J_{\mu,i}$ generators of Eq.~(\ref{jmui})
we obtain the ${\cal T}_d$ tensors
\ba
\hat T^{\Gamma}_{\mu,\gamma} &=& 
\sum_{i=1}^{6} \, \alpha^{\Gamma}_{\gamma,i} \, \hat J_{\mu,i} ~,
\label{alpha}
\ea
where $\mu=\pm,0$ and $\gamma$ denotes the component of $\Gamma$. 
The explicit expressions  are given by
\ba
\hat T^{A_1}_{\mu,1} &=& \frac{1}{\sqrt{6}}  
\sum_{i=1}^{6} \, \hat J_{\mu,i} ~, 
\nonumber\\ 
\hat T^{E}_{\mu,1} &=& \frac{1}{2\sqrt{3}} \left( \hat J_{\mu,1} 
+ \hat J_{\mu,2} - 2 \hat J_{\mu,3} + \hat J_{\mu,4} 
- 2 \hat J_{\mu,5} + \hat J_{\mu,6} \right) ~, 
\nonumber\\ 
\hat T^{E}_{\mu,2} &=& \frac{1}{2} \left( \hat J_{\mu,1} 
- \hat J_{\mu,2} - \hat J_{\mu,4} + \hat J_{\mu,6} \right) ~, 
\nonumber\\
\hat T^{F_2}_{\mu,1} &=& \frac{1}{\sqrt{2}} 
\left( \hat J_{\mu,1} - \hat J_{\mu,6} \right) ~,
\nonumber\\
\hat T^{F_2}_{\mu,2} &=& \frac{1}{\sqrt{2}} 
\left( \hat J_{\mu,2} - \hat J_{\mu,4} \right) ~,
\nonumber\\
\hat T^{F_2}_{\mu,3} &=& \frac{1}{\sqrt{2}} 
\left( \hat J_{\mu,3} - \hat J_{\mu,5} \right) ~. \label{tdgen}
\ea
The algebraic Hamiltonian can now be constructed by repeated couplings 
of these tensors to a total symmetry $A_1$, since it must commute 
with all operations in ${\cal T}_d$. This is accomplished by means
of the ${\cal T}_d$ Clebsch-Gordan coefficients \cite{LF,bible}. 

In order to establish a connection with configuration space calculations 
we use the analysis of \cite{Hecht}. In this work the vibrational 
Hamiltonian for the Be$_4$ cluster is expressed in terms of 
symmetry-adapted internal coordinates, 
$q^{\Gamma}_{\gamma}$, and momenta, $p^{\Gamma}_{\gamma}$. 
The transformation to the tensor operators of 
Eqs.~(\ref{alpha},\ref{tdgen}) 
proceeds in two steps. First we introduce creation and annihilation 
operators 
\ba
b^{\Gamma \, \dagger}_{\gamma} \;=\; \frac{1}{\sqrt{2}} \left( 
q^{\Gamma}_{\gamma} - i p^{\Gamma}_{\gamma} \right) ~, \hspace{1cm}
b^{\Gamma}_{\gamma} \;=\; \frac{1}{\sqrt{2}} \left( 
q^{\Gamma}_{\gamma} + i p^{\Gamma}_{\gamma} \right) ~. 
\ea
Here the normal boson operators are related to the local boson operators 
by $b^{\Gamma}_{\gamma} = \sum_{i} \alpha^{\Gamma}_{\gamma,i} \, b_i$. 
The coefficients $\alpha^{\Gamma}_{\gamma \, i}$ can be read from 
Eqs.~(\ref{alpha},\ref{tdgen}). Next the local boson operators are 
associated with the generators of Eq.~(\ref{jmui}) by means of 
\ba
b_i \rightarrow \hat J_{+,i}/\sqrt{N_i} ~, \hspace{1cm} 
b^{\dagger}_i \rightarrow \hat J_{-,i}/\sqrt{N_i} ~.
\label{subst}
\ea
This transformation is such that the commutator 
\ba
\frac{1}{N_i} [ \hat J_{+,i},\hat J_{-,i}] 
\;=\; \frac{1}{N_i} 2\hat J_{0,i} \
\;=\; \frac{1}{N_i} (\hat N_i - 2 \hat v_i)
\;=\; 1 - \frac{2 \hat v_i}{N_i} ~, \label{comm}
\ea
reduces for $N_i \rightarrow \infty$ to the standard boson commutator 
$[b_i,b^\dagger_i]=1$.  
Eq.~(\ref{subst}) provides a procedure to 
construct an anharmonic representation of harmonic operators. 
The anharmonic contribution arises from the $-2\hat v_i/N_i$ term 
in Eq.~(\ref{comm}). Note that each local harmonic 
oscillator $(b^{\dagger}_i b_i + b_i b^{\dagger}_i)/2$ leads to 
$(\hat J_{-,i} \hat J_{+,i} + \hat J_{+,i} \hat J_{-,i})/2N_i 
=\hat v_i + 1/2 - \hat v_i^2/N_i$ and thus to a Morse-like spectrum 
through the association of Eq.~(\ref{subst}). In this way the 
algebraic model substitutes harmonic oscillators by anharmonic ones. 

This procedure can be applied to the various contributions to the 
vibrational Hamiltonian for the Be$_4$ cluster \cite{Hecht}.
The vibrational basis states for this system are usually labeled by 
$(\nu_1,\nu_2^m,\nu_3^l)$ \cite{Herzberg}. 
Here $\nu_1$, $\nu_2$ and $\nu_3$ denote 
the number of phonons in the $A_1$, $E$ and $F_2$ modes, respectively, 
and $m=\nu_2,\nu_2-2, \ldots, 1$ (or 0) for $\nu_2$ odd (or even) and 
and $l=\nu_3,\nu_3-2, \ldots, 1$ (or 0) for $\nu_3$ odd (or even)
are the vibrational angular momenta associated with the 
$E$ and $F_2$ modes. 
For the zeroth order vibrational Hamiltonian we find 
\ba
H_0 &=& \omega_1 \, \hat{\cal H}_{A_1} + \omega_2 \, \hat{\cal H}_{E}
+ \omega_3 \, \hat{\cal H}_{F_2} ~, \label{h0}
\ea
with
\ba
\hat{\cal H}_{\Gamma} &=& \frac{1}{2N} \sum_{\gamma} \left( 
  \hat T^{\Gamma}_{-,\gamma} \hat T^{\Gamma}_{+,\gamma}
+ \hat T^{\Gamma}_{+,\gamma} \hat T^{\Gamma}_{-,\gamma} \right) ~. 
\label{hgamma}
\ea
The anharmonic vibrational terms are expressed in terms of products 
of $\hat{\cal H}_{\Gamma}$,
\ba
H_1 &=& X_{11} \left( \hat{\cal H}_{A_1} \right)^2
      + X_{22} \left( \hat{\cal H}_{E}   \right)^2
      + X_{33} \left( \hat{\cal H}_{F_2} \right)^2
\nonumber\\
&&    + X_{12} \left( \hat{\cal H}_{A_1} \hat{\cal H}_{E  } \right)
      + X_{13} \left( \hat{\cal H}_{A_1} \hat{\cal H}_{F_2} \right)
      + X_{23} \left( \hat{\cal H}_{E  } \hat{\cal H}_{F_2} \right) ~.
\label{h1}
\ea
The further splitting of vibrational levels $(\nu_1,\nu_2,\nu_3)$ 
into its possible sublevels is achieved by means of the interactions 
\cite{Hecht} 
\ba
H_2 &=& g_{22} \, \left( \hat l^{A_2} \right)^2  
+ g_{33} \, \sum_{\gamma} \hat l^{F_1}_{\gamma} \, \hat
l^{F_1}_{\gamma} 
\nonumber\\
&& + t_{33} \left( 6 \sum_{\gamma} 
[ \hat T^{F_2}_- \times \hat T^{F_2}_- ]^{E}_{\gamma} \,
[ \hat T^{F_2}_+ \times \hat T^{F_2}_+ ]^{E}_{\gamma} 
-4 \sum_{\gamma}
[ \hat T^{F_2}_- \times \hat T^{F_2}_- ]^{F_2}_{\gamma} \,
[ \hat T^{F_2}_+ \times \hat T^{F_2}_+ ]^{F_2}_{\gamma} \right) 
\frac{1}{N^2}
\nonumber\\
&& + t_{23} \left( 8 \sum_{\gamma} 
[ \hat T^{E}_- \times \hat T^{F_2}_- ]^{F_1}_{\gamma} \,
[ \hat T^{E}_+ \times \hat T^{F_2}_+ ]^{F_1}_{\gamma} 
-8 \sum_{\gamma}
[ \hat T^{E}_- \times \hat T^{F_2}_- ]^{F_2}_{\gamma} \,
[ \hat T^{E}_+ \times \hat T^{F_2}_+ ]^{F_2}_{\gamma} \right) 
\frac{1}{N^2} ~.
\label{h2}
\ea
The operators $\hat l^{A_2}$ and $\hat l^{F_1}$ represent the vibrational 
angular momentum operators associated with the $E$ and $F_2$ modes, 
respectively,
\ba 
\hat l^{A_2} &=& -i \, \sqrt{2} \frac{1}{N} 
[ \hat T^E_{-} \times \hat T^E_{+} ]^{A_2} ~,
\nonumber\\
\hat l^{F_1}_{\gamma} &=& +i \, \sqrt{2} \frac{1}{N} 
[ \hat T^{F_2}_{-} \times \hat T^{F_2}_{+} ]^{F_1}_{\gamma} ~.
\label{vibang}
\ea
The square brackets in Eqs.~(\ref{h2},\ref{vibang}) 
denote the tensor coupling under the point group ${\cal T}_d$
\ba
[ \hat T^{\Gamma_1} \times \hat T^{\Gamma_2} ]^{\Gamma}_{\gamma} 
&=& \sum_{\gamma_1,\gamma_2} 
C(\Gamma_1,\Gamma_2,\Gamma ; \gamma_1,\gamma_2,\gamma) \,
\hat T^{\Gamma_1}_{\gamma_1} \, \hat T^{\Gamma_2}_{\gamma_2} ~,
\label{tensor}
\ea
where the expansion coefficients are the Clebsch-Gordan coefficients for 
${\cal T}_d$ \cite{LF,bible}. The interactions of Eq.~(\ref{h2}) were 
absent in previous versions of the model \cite{vibron1,vibron2,IOFL}.

The algebraic Hamiltonian of Eqs~(\ref{h0}--\ref{tensor}) is the 
algebraic equivalent of the vibrational Hamiltonian of \cite{Hecht}.
The harmonic frequencies $\omega_i$ and anharmonic constants 
$X_{ij}$, $g_{22}$, $g_{33}$, $t_{33}$ and $t_{23}$ have the same 
meaning as in \cite{Hecht} and the ab initio calculations of \cite{RLT} 
can be used to generate the spectrum.
The various contributions to the algebraic Hamiltonian arise
naturally from the successive couplings of the fundamental tensors 
of Eq.~(\ref{tdgen}). 
The scale transformation of Eq.~(\ref{subst}) makes it possible to 
establish the connection between ab initio and algebraic 
parameteres and to explicitly construct the algebraic interactions 
that correspond to interactions in configuration space. 
In the opposite sense, Eqs.~(\ref{subst},\ref{comm}) provide a procedure 
to obtain a geometric interpretation of algebraic interactions in terms of 
those in configuration space. In the harmonic limit, which is defined as 
$N_i \rightarrow \infty$, Eq.~(\ref{comm}) reduces to the standard 
boson commutator $[b_i,b_i^{\dagger}]=1$. This limit corresponds to a 
contraction of $SU_i(2)$ to the Weyl algebra, generated by the set 
$\{ b_i,b_i^{\dagger},1 \}$.
In the harmonic limit the Hamiltonian $H_0+H_1+H_2$ of 
Eqs.~(\ref{h0}--\ref{tensor}) reduces {\em exactly} to the vibrational 
Hamiltonian of \cite{Hecht}. Note that because of the replacement of 
Eq.~(\ref{subst}) the Hamiltonian $H_0+H_1+H_2$ only depends on the 
$\hat T^{\Gamma}_{\mu,\gamma}$ tensors of Eq.~(\ref{tdgen}) with 
$\mu=\pm$. In addition, the algebraic model provides terms involving the 
$\hat T^{\Gamma}_{\mu,\gamma}$ tensors with $\mu=0$. 
As can be seen from Eq.~(\ref{comm}) these terms are completely 
anharmonic in origin and have no direct counterpart in  models based
on the standard harmonic bosons. These operators arise from the 
substitution of harmonic oscillators by Morse oscillators and play 
an important role when dealing with anharmonic molecules, particularly 
at higher phonon numbers \cite{nosotros}.

Apart from providing a direct connection to configuration space 
calculations, this formalism can also be used as an effective 
model of molecular vibrations, particularly when no ab initio 
calculations are available. As an example, we show in Table~\ref{Be4} 
the results of a fit to the ab initio calculations for Be$_4$ up to 
four phonons. The ab initio results were generated with the parameters 
from \cite{RLT}. The Hamiltonian used in the fit contains 9 interaction 
terms compared to the 13 of \cite{Hecht} 
(see Eqs.~(\ref{h0},\ref{h1},\ref{h2})).  
The parameters are extracted in a fit that includes all vibrational 
energies up to four phonons ($V=\nu_1+\nu_2+\nu_3 \leq 4$):
$\omega_1=636$, $\omega_2=453$, $\omega_3=532$, 
$X_{33}=44.276$, $X_{12}= 4.546$, $X_{13}=-2.539$, 
$g_{33}=-15.031$, $t_{33}=-1.679$ and $t_{23}=-1.175$. 
All values are given in cm$^{-1}$. The total number of bosons 
used in the fit is $N=44$. The r.m.s. deviation between 
the algebraic and the ab initio calculations is 2.6 cm$^{-1}$. 

As a test of the predictive power of the algebraic approach we have 
performed another calculation in which the same 9 parameters were determined 
in a fit that only included the vibrational energies up to three phonons 
($V \leq 3$). In this case the r.m.s. deviation is 1.6 cm$^{-1}$. 
If we now use these values of the parameters to calculate the four phonon 
states, the r.m.s. increases to only 3.1 cm$^{-1}$, compared to 
2.6 cm$^{-1}$ in the previous calculation. We remark that by restricting 
the model interactions to Casimir invariants and their powers 
\cite{vibron1,vibron2,IOFL} the Be$_4$ spectrum cannot be reproduced. 
The terms in $H_2$ of Eq.~(\ref{h2}) play a crucial role. 

Repeating the same fit in the harmonic limit ($N \rightarrow \infty$) the 
r.m.s. deviation increases from 2.6 to 5.4 cm$^{-1}$. This shows that the 
anharmonic contributions introduced by taking a finite value of $N$ 
(see Eq.~(\ref{comm})) provide an important improvement of the fit. 
The real test of this aspect is a fit to experimental data rather than 
to other calculations. 
Work on the application of this algebraic model to the experimental 
vibrational spectra of polyatomic molecules is in progress 
\cite{nosotros}.

In summary, in this letter we have established a 
connection between algebraic and configuration-space interactions. 
For the example of the Be$_4$ cluster (with tetrahedral symmetry) 
we have, starting from configuration space interactions, constructed 
explicitly the corresponding algebraic interactions (which have a 
richer structure than in previous versions of the model). In the harmonic 
limit the configuration space results are reproduced exactly. 

In addition, it was shown that the algebraic model can also be used 
as an effective model of molecular vibrations with good precision. 
In the algebraic approach the eigenvalues and corresponding wave 
functions are obtained by matrix diagonalization. Hence the required 
computing time is small. These properties open the possibility to 
use the algebraic model as a numerically efficient, empirical tool 
to study molecular vibrations, especially when no ab initio calculations 
are available (or feasible) \cite{nosotros}. 

We thank F. Iachello and P. van Isacker for their continuous
interest and useful comments. This work was supported in part by the 
European Community under contract nr. CI1$^{\ast}$-CT94-0072, 
DGAPA-UNAM under project IN105194, CONACyT-M\'exico under project 
400340-5-3401E and Spanish DGCYT under project PB92-0663.

\begin{table}
\centering
\caption[]{\small
Vibrational excitations of Be$_4$ using the algebraic Hamiltonian 
with parameters given in the text. The ab initio 
($N \rightarrow \infty$) spectrum is generated with the parameters 
from \cite{RLT}. The energies are given in cm$^{-1}$.
\normalsize}
\vspace{10pt} \label{Be4}
\begin{tabular}{cclcc|cclcc}
\hline
& & & & & & & & & \\
$V$ & $(\nu_1,\nu_2^m,\nu_3^l)$ & $\Gamma$ 
& Ab initio & Fit &
$V$ & $(\nu_1,\nu_2^m,\nu_3^l)$ & $\Gamma$ 
& Ab initio & Fit \\
& & & $N \rightarrow \infty$ & $N=44$ & 
& & & $N \rightarrow \infty$ & $N=44$ \\
& & & & & & & & & \\
\hline
& & & & \\
1 & $(1,0^0,0^0)$     & $A_1$ &  638.6 &  637.0 & 
3 & $(1,0^0,2^0)$     & $A_1$ & 2106.8 & 2105.6 \\
  & $(0,1^1,0^0)$     & $E$   &  453.6 &  455.0 & 
  & $(1,0^0,2^2)$     & $E$   & 2000.1 & 1999.8 \\
  & $(0,0^0,1^1)$     & $F_2$ &  681.9 &  678.2 & 
  &                   & $F_2$ & 2056.8 & 2052.8 \\
2 & $(2,0^0,0^0)$     & $A_1$ & 1271.0 & 1269.2 & 
  & $(0,3^1,0^0)$     & $E$   & 1341.3 & 1343.7 \\
  & $(1,1^1,0^0)$     & $E$   & 1087.1 & 1087.0 & 
  & $(0,3^3,0^0)$     & $A_1$ & 1355.5 & 1352.5 \\
  & $(1,0^0,1^1)$     & $F_2$ & 1312.6 & 1308.3 & 
  &                   & $A_2$ & 1355.5 & 1354.4 \\
  & $(0,2^0,0^0)$     & $A_1$ &  898.3 &  901.4 & 
  & $(0,2^{0,2},1^1)$ & $F_2$ & 1565.5 & 1565.7 \\
  & $(0,2^2,0^0)$     & $E$   &  905.4 &  906.1 & 
  &                   & $F_2$ & 1584.4 & 1583.1 \\
  & $(0,1^1,1^1)$     & $F_1$ & 1126.7 & 1125.1 & 
  & $(0,2^2,1^1)$     & $F_1$ & 1578.5 & 1578.0 \\
  &                   & $F_2$ & 1135.5 & 1134.1 & 
  & $(0,1^1,2^{0,2})$ & $E$   & 1821.4 & 1821.6 \\
  & $(0,0^0,2^0)$     & $A_1$ & 1484.0 & 1483.0 & 
  &                   & $E$   & 1929.5 & 1929.0 \\
  & $(0,0^0,2^2)$     & $E$   & 1377.3 & 1373.9 & 
  & $(0,1^1,2^2)$     & $A_2$ & 1813.3 & 1813.1 \\
  &                   & $F_2$ & 1434.1 & 1429.6 & 
  &                   & $A_1$ & 1830.8 & 1831.7 \\
3 & $(3,0^0,0^0)$     & $A_1$ & 1897.0 & 1896.7 & 
  &                   & $F_2$ & 1874.4 & 1873.2 \\
  & $(2,1^1,0^0)$     & $E$   & 1714.3 & 1714.3 & 
  &                   & $F_1$ & 1883.2 & 1883.0 \\
  & $(2,0^0,1^1)$     & $F_2$ & 1937.0 & 1933.7 & 
  & $(0,0^0,3^{1,3})$ & $F_2$ & 2136.5 & 2134.2 \\
  & $(1,2^0,0^0)$     & $A_1$ & 1526.6 & 1529.2 & 
  &                   & $F_2$ & 2327.3 & 2326.9 \\
  & $(1,2^2,0^0)$     & $E$   & 1533.7 & 1532.8 & 
  & $(0,0^0,3^3)$     & $F_1$ & 2199.8 & 2197.1 \\
  & $(1,1^1,1^1)$     & $F_1$ & 1752.2 & 1749.7 & 
  &                   & $A_1$ & 2256.5 & 2254.4 \\
  &                   & $F_2$ & 1761.0 & 1759.8 & 
  &                   &       &        &        \\
& & & & & & & & & \\
\hline
\end{tabular}
\end{table}

\addtocounter{table}{-1}
\clearpage
\begin{table}
\centering
\caption[]{\small Continued. \normalsize}
\vspace{10pt}
\begin{tabular}{cclcc|cclcc}
\hline
& & & & & & & & & \\
$V$ & $(\nu_1,\nu_2^m,\nu_3^l)$ & $\Gamma$ 
& Ab initio & Fit &
$V$ & $(\nu_1,\nu_2^m,\nu_3^l)$ & $\Gamma$ 
& Ab initio & Fit \\
& & & $N \rightarrow \infty$ & $N=44$ & 
& & & $N \rightarrow \infty$ & $N=44$ \\
& & & & & & & & & \\
\hline
& & & & & & & & & \\
4 & $(4,0^0,0^0)$         & $A_1$ & 2516.8 & 2519.5 & 
4 & $(0,4^4,0^0)$         & $E$   & 1803.8 & 1797.1 \\
  & $(3,1^1,0^0)$         & $E$   & 2335.2 & 2336.9 & 
  & $(0,3^{1,3},1^1)$     & $F_1$ & 1998.9 & 2000.1 \\
  & $(3,0^0,1^1)$         & $F_2$ & 2555.1 & 2554.4 & 
  &                       & $F_2$ & 2013.3 & 2014.0 \\
  & $(2,2^0,0^0)$         & $A_1$ & 2148.7 & 2152.2 & 
  &                       & $F_1$ & 2026.4 & 2025.0 \\
  & $(2,2^2,0^0)$         & $E$   & 2155.8 & 2154.8 & 
  &                       & $F_2$ & 2029.5 & 2024.7 \\
  & $(2,1^1,1^1)$         & $F_1$ & 2371.5 & 2369.8 & 
  & $(0,2^{0,2},2^{0,2})$ & $E$   & 2247.8 & 2251.0 \\
  &                       & $F_2$ & 2380.2 & 2380.7 & 
  &                       & $A_1$ & 2262.1 & 2263.0 \\ 
  & $(2,0^0,2^0)$         & $A_1$ & 2723.2 & 2723.6 & 
  &                       & $E$   & 2273.9 & 2276.6 \\ 
  & $(2,0^0,2^2)$         & $E$   & 2616.5 & 2620.8 & 
  &                       & $A_1$ & 2367.6 & 2367.0 \\
  &                       & $F_2$ & 2673.3 & 2671.4 & 
  &                       & $E$   & 2373.1 & 2371.1 \\   
  & $(1,3^1,0^0)$         & $E$   & 1964.4 & 1967.1 & 
  & $(0,2^{0,2},2^2)$     & $F_2$ & 2308.8 & 2310.9 \\
  & $(1,3^3,0^0)$         & $A_1$ & 1978.7 & 1973.6 &
  &                       & $F_2$ & 2327.7 & 2330.2 \\
  &                       & $A_2$ & 1978.7 & 1975.2 &  
  & $(0,2^2,2^2)$         & $A_2$ & 2265.1 & 2268.2 \\
  & $(1,2^{0,2},1^1)$     & $F_2$ & 2185.8 & 2185.3 &
  &                       & $F_1$ & 2321.8 & 2321.9 \\
  &                       & $F_2$ & 2204.8 & 2204.5 &  
  & $(0,1^1,3^{1,3})$     & $F_1$ & 2567.1 & 2570.0 \\
  & $(1,2^2,1^1)$         & $F_1$ & 2198.9 & 2197.7 & 
  &                       & $F_2$ & 2585.5 & 2588.3 \\
  & $(1,1^1,2^{0,2})$     & $E$   & 2438.9 & 2443.5 &
  &                       & $F_1$ & 2639.9 & 2643.1 \\
  &                       & $E$   & 2547.1 & 2545.7 &  
  &                       & $F_2$ & 2640.1 & 2643.0 \\
  & $(1,1^1,2^2)$         & $A_2$ & 2430.9 & 2431.8 & 
  &                       & $F_1$ & 2764.3 & 2764.9 \\
  &                       & $A_1$ & 2448.4 & 2455.1 &
  &                       & $F_2$ & 2772.0 & 2779.7 \\   
  &                       & $F_2$ & 2492.0 & 2491.0 & 
  & $(0,1^1,3^3)$         & $E$   & 2696.8 & 2700.3 \\
  &                       & $F_1$ & 2500.8 & 2501.5 &  
  & $(0,0^0,4^{0,4})$     & $A_1$ & 2909.1 & 2906.1 \\
  & $(1,0^0,3^{1,3})$     & $F_2$ & 2751.2 & 2748.4 &
  &                       & $A_1$ & 3290.9 & 3290.5 \\ 
  &                       & $F_2$ & 2942.1 & 2942.6 &  
  & $(0,0^0,4^{2,4})$     & $E$   & 2956.1 & 2952.6 \\
  & $(1,0^0,3^3)$         & $F_1$ & 2814.5 & 2816.6 &
  &                       & $F_2$ & 3067.3 & 3067.3 \\
  &                       & $A_1$ & 2871.3 & 2870.7 &  
  &                       & $E$   & 3137.2 & 3134.9 \\
  & $(0,4^0,0^0)$         & $A_1$ & 1775.3 & 1776.8 & 
  &                       & $F_2$ & 3253.0 & 3253.7 \\
  & $(0,4^2,0^0)$         & $E$   & 1782.5 & 1781.6 & 
  & $(0,0^0,4^4)$         & $F_1$ & 2978.9 & 2978.0 \\
& & & & & & & & & \\
\hline
\end{tabular}
\end{table}

\end{document}